\documentclass[a4paper,11pt]{article}
\usepackage{pos}
\usepackage{physics}

\title{Unified thermal model for photohadronic neutrino production in astrophysical sources}
\author[a,b]{Damiano F.G. Fiorillo}

\affiliation[a]{Dipartimento di Fisica "Ettore Pancini", Universit\'a degli studi di Napoli "Federico II", Complesso Univ. Monte S. Angelo, I-80126 Napoli, Italy}

\emailAdd{dfgfiorillo@na.infn.it}

\abstract{Astrophysical neutrino fluxes are often modeled as power laws of the energy. This is reasonable in the case of hadronic sources, but it does not capture the behavior in photohadronic sources, where the spectrum depends on the properties of the target photons on which protons collide. This limits the possibility of a unified treatment of different sources. In order to overcome this difficulty, we model the target photons by a blackbody spectrum. This model is sufficiently flexible to reproduce neutrino fluxes from known photohadronic sources; we apply it to study the sensitivity of Dense Neutrino Arrays, Neutrino Telescopes and Neutrino Radio Arrays to photohadronic sources. We also classify the flavor composition of the neutrino spectrum in terms of the parameter space. We discuss the interplay with the experiments, studying the changes in the track-to-shower ratio induced by different flavor compositions, both within and outside the region of the Glashow resonance.}

\FullConference{37$^{\rm{th}}$ International Cosmic Ray Conference (ICRC 2021)\\
		July 12th -- 23rd, 2021\\
		Online -- Berlin, Germany}

\begin{document}
\maketitle

\section{Introduction}

High-energy neutrinos can be produced in astrophysical sources either in proton-proton collisions of cosmic-rays with gas or in proton-photon, or photohadronic, collisions of cosmic-rays with the radiation field in the source. In the first case, the neutrino spectrum follows the parent cosmic-ray spectrum, and therefore a simple parameterization is provided by the simple power law approximation. However, for neutrinos produced in photohadronic collisions, the neutrino spectrum depends both on the cosmic-ray spectrum and on the spectrum of the low-energy photons acting as a target for $p\gamma$ collisions. The target photon spectrum can influence the detection prospects, since it determines the energy range in which the neutrino spectrum will be peaked. Furthermore, it can influence the flavor composition at different energies and the relative number of neutrinos and antineutrinos. Since target photons can be of very different nature among different source classes, typically photohadronic neutrinos are obtained by dedicated modeling for each class, e.g. for Active Galactic Nuclei (AGN), Gamma-Ray Bursts (GRBs), and Tidal Disruption Events (TDEs). This approach is clearly unsuited for systematic studies aiming to describe different source classes within a common framework.

Here we propose a model which can describe photohadronic production in astrophysical sources in a unified way. The main idea of the model is to replace the real target photon spectrum by a fictitious blackbody spectrum, whose temperature is suitably chosen to reproduce approximately the neutrino spectrum. While we refer to Ref.~\cite{Fiorillo:2021hty} for a detailed treatment of the model, we test it on benchmark astrophysical sources with non-thermal target photon spectrum to verify that the neutrino production is correctly reproduced. We then apply the model in two different directions: first of all, we study the sensitivity of neutrino detectors sensitive in different energy ranges to neutrinos from astrophysical sources. In this way, we are able to identify which detector is more suitable for neutrino detection from a given source. As a second application, we determine the flavor composition as a function of the energy and we discuss how it is influenced by the parameters of the astrophysical source.

\section{Thermal model}

The spectral shape of neutrinos produced by $p\gamma$ collisions depend both on the cosmic-ray and on the target photon spectrum. For cosmic-rays the spectral shape is commonly assumed to be a power law, as expected if they are accelerated by Fermi mechanism, with a maximal proton energy. The latter is determined either by the strongest of two conditions: that cosmic-rays are confined within the source by its magnetic field, leading to the well-known Hillas criterion, and that the synchrotron energy losses in the magnetic field do not exceed the rate of energy increase via acceleration. However, for target photons such a general assumption is not possible, since their origin could be due to entirely different processes for different astrophysical sources: for example, for GRBs and AGN the target photons are generally described by broken power laws~\cite{Baerwald:2011ee,Hummer:2010ai,Gao:2016uld}, whereas for TDEs a thermal shape for the target photons has been assumed in the literature~\cite{Winter:2020ptf}. This large variability is the main difficulty in capturing the variety of $p\gamma$ sources with a single model.

In order to overcome this difficulty, we identify a single photon energy $\bar{\varepsilon}'_\gamma$ contributing most to $p\gamma$ production, and we replace the real target photon spectrum with a fictitious blackbody spectrum peaked at this energy $\bar{\varepsilon}'_\gamma$. Therefore, we provide a simple mapping between the real photon spectrum, the energy $\bar{\varepsilon}'_\gamma$, and the corresponding blackbody spectrum. Let us first of all discuss how the energy $\bar{\varepsilon}'_\gamma$ can be identified. In all this discussion we will use primed quantities referred to the frame comoving with the source: since the latter could be subject to relativistic motion, such as jet expansion, we will denote the quantities in the observer frame as unprimed. The energies in the two frames are related by $E=\Gamma E'$, where $\Gamma$ is the Doppler factor.

Let the number of target photons per unit volume per unit energy be $n(\varepsilon'_\gamma)$. The interaction rate of a proton with this photon target is then proportional to the cross section for photohadronic interaction and to the photon {\em number} density $\varepsilon'_\gamma n(\varepsilon'_\gamma)$. The  interaction rate vanishes if the photon energy is too small (below threshold), because of the vanishing of the cross section. For a proton with energy $E'$, with good approximation the photons that participate to $p\gamma$ interactions have energies $\varepsilon'_\gamma\gtrsim y_\Delta m_p/E'$, where $y_\Delta \simeq 0.2$~GeV. Therefore, for a distribution of protons with a maximal energy $E'_\text{p,max}$, only those photons are important for $p\gamma$ interaction which satisfy the condition
\begin{equation}\label{eq:kincond}
    \varepsilon'_\gamma>\frac{y_\Delta m_p}{E'_\text{p,max}}.
\end{equation}
The energy $\bar{\varepsilon}'_\gamma$ must of course be sought for in this range. Since the interaction rate is proportional to the number density $\varepsilon'_\gamma n(\varepsilon'_\gamma)$, we expect that the energy contributing most to $p\gamma$ interaction is the energy maximizing the number density. Therefore we define $\bar{\varepsilon}'_\gamma$ as the energy at which $\varepsilon'_\gamma n(\varepsilon'_\gamma)$ is maximum. This definition is general enough to capture most of the cases. 

\begin{figure}[t!]
    \centering
    \includegraphics[width=0.445\textwidth]{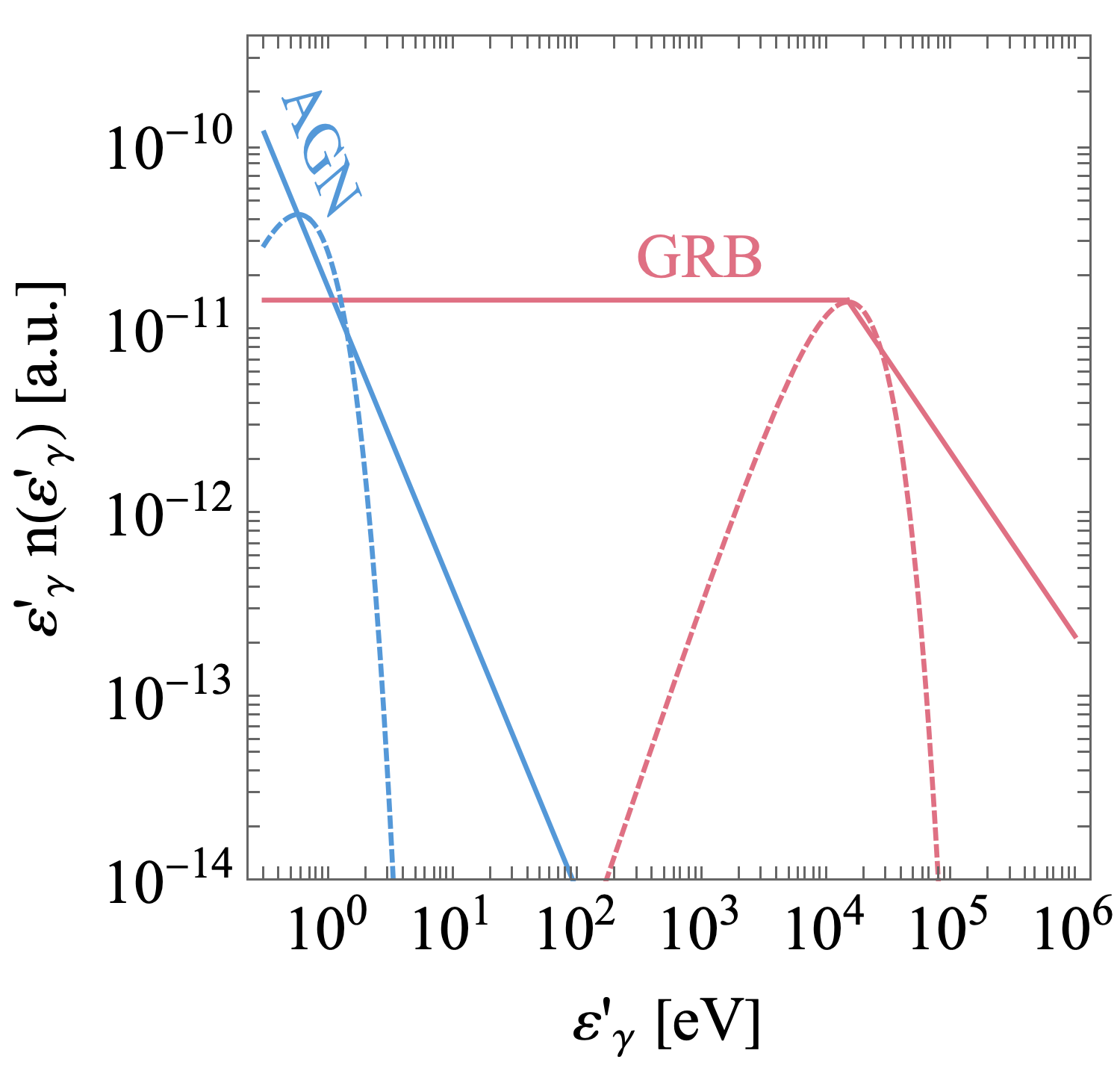}
    \includegraphics[width=0.45\textwidth]{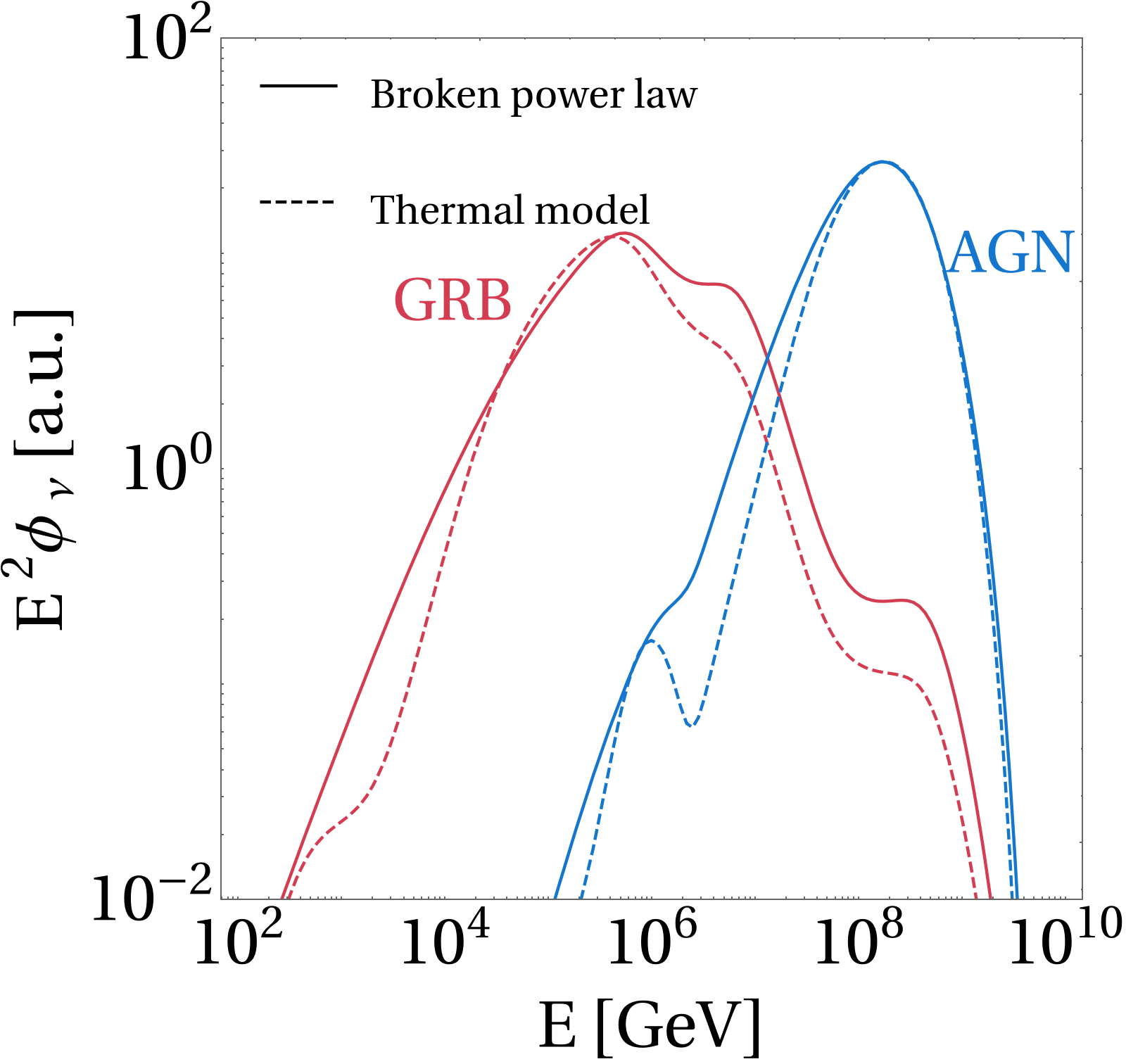}
    \caption{Comparison between the benchmark astrophysical neutrino fluxes, parameterized as broken power laws, and their reproduction using the thermal model. In the left panel we show the target photon spectra $\varepsilon'_\gamma n(\varepsilon'_\gamma)$ (normalization in arbitrary units). In the right panel we show the comparison for the neutrino fluxes produced in AGN and GRBs. We show the all-flavor neutrino flux $E^2 \phi$ in GeV~cm$^{-2}$ s$^{-1}$ (normalization in arbitrary units) as a function of the energy in the observer rest frame. In the right panel we show the fraction of electron neutrinos and antineutrinos in the differential flux at the source as a function of the energy in the observer rest frame, in the energy region in which the flux is at least $1/1000$ of its peak value. The solid curves are obtained from the astrophysical broken power-law source model, the dashed curves are obtained with the thermal model. }
    \label{thecomparison}
\end{figure}

One situation which needs more specification is the case in which the number density is flat over an energy interval, which happens if $n(\varepsilon'_\gamma)\propto \varepsilon^{'-1}_\gamma$. This happens, for example, in GRBs, where the prompt emission of photons often exhibit a flat number density of photons in energy below tens or hundreds of keV. In this situation, a single maximum cannot be identified: as shown in Ref.~\cite{Fiorillo:2021hty}, the best choice for $\bar{\varepsilon}'_\gamma$ is now the upper bound of the interval over which $\varepsilon'_\gamma n(\varepsilon'_\gamma)$ is flat in energy. For example, for the case of GRBs, a reasonable photon distribution can be parameterized as~\cite{Baerwald:2011ee}
\begin{align} \label{eq:brokenpower}
    n_\gamma(\varepsilon'_\gamma)\propto\begin{cases} (\varepsilon'_\gamma/\varepsilon'_\text{b})^{ -\alpha}, \hspace{0.2cm} & \varepsilon'_{\text{min}} \leq \varepsilon'_\gamma\leq\varepsilon'_{\text{b}} \\ (\varepsilon'_\gamma/\varepsilon'_\text{b})^{-\beta},\hspace{0.2 cm} & \varepsilon'_{\text{b}} < \varepsilon'_\gamma\leq \varepsilon'_{\text{max}} \end{cases}
\end{align}
where $\alpha=1$ and $\beta=2$. The values of the energies $\varepsilon'_{\text{min}}$, $\varepsilon'_{\text{b}}$, $\varepsilon'_{\text{max}}$ are reported in Ref.~\cite{Fiorillo:2021hty}. The number density $\varepsilon'_\gamma n(\varepsilon'_\gamma)$ is flat for energies below $\varepsilon'_{\text{b}}$: according to our criterion above, we would then choose $\bar{\varepsilon}'_{\gamma}=\varepsilon'_{\text{b}}$.

Having determined the map between a generic target photon spectrum and the energy $\bar{\varepsilon}'_\gamma$, we can now replace the photon spectrum by a fictitious blackbody spectrum $n^{\text{th}}(\varepsilon'_\gamma)$ with an effective temperature $T'$. The blackbody number density $\varepsilon'_\gamma n^{\text{th}}(\varepsilon'_\gamma)$ is peaked at an energy $\varepsilon'_\gamma=2.8T'$: by requiring this to be equal to $\bar{\varepsilon}'_\gamma$, we obtain the effective temperature
\begin{equation}
T'=\frac{\bar{\varepsilon}'_\gamma}{2.8}.
\end{equation}
In order to show the goodness in the reproduction, we take two benchmark sources as an example, namely AGN and GRBs. Both these sources have a target photon spectrum that is well approximated by a broken power law (only in the region most relevant for neutrino production). The parameters of this power law are reported in Ref.~\cite{Fiorillo:2021hty}, as well as the effective temperatures of the blackbody spectrum obtained from the procedure above. The determination of the neutrino spectrum is done using the software NeuCosmA~\cite{Hummer:2010ai}. Using as input the magnetic field of the source $B'$, the source size $R'$, the Doppler factor $\Gamma$, and the target photon spectrum, NeuCosmA determines the produced neutrino spectrum, accounting also for the cooling of the secondary particles due to synchrotron losses in the source magnetic field. The comparison between the neutrino spectrum using the broken power law for the photons (solid) and using the thermal model (dashed) shows that the latter is able to reproduce quite accurately the neutrino production from the sources. 

\begin{figure}[t!]
    \centering
    \includegraphics[width=0.40\textwidth]{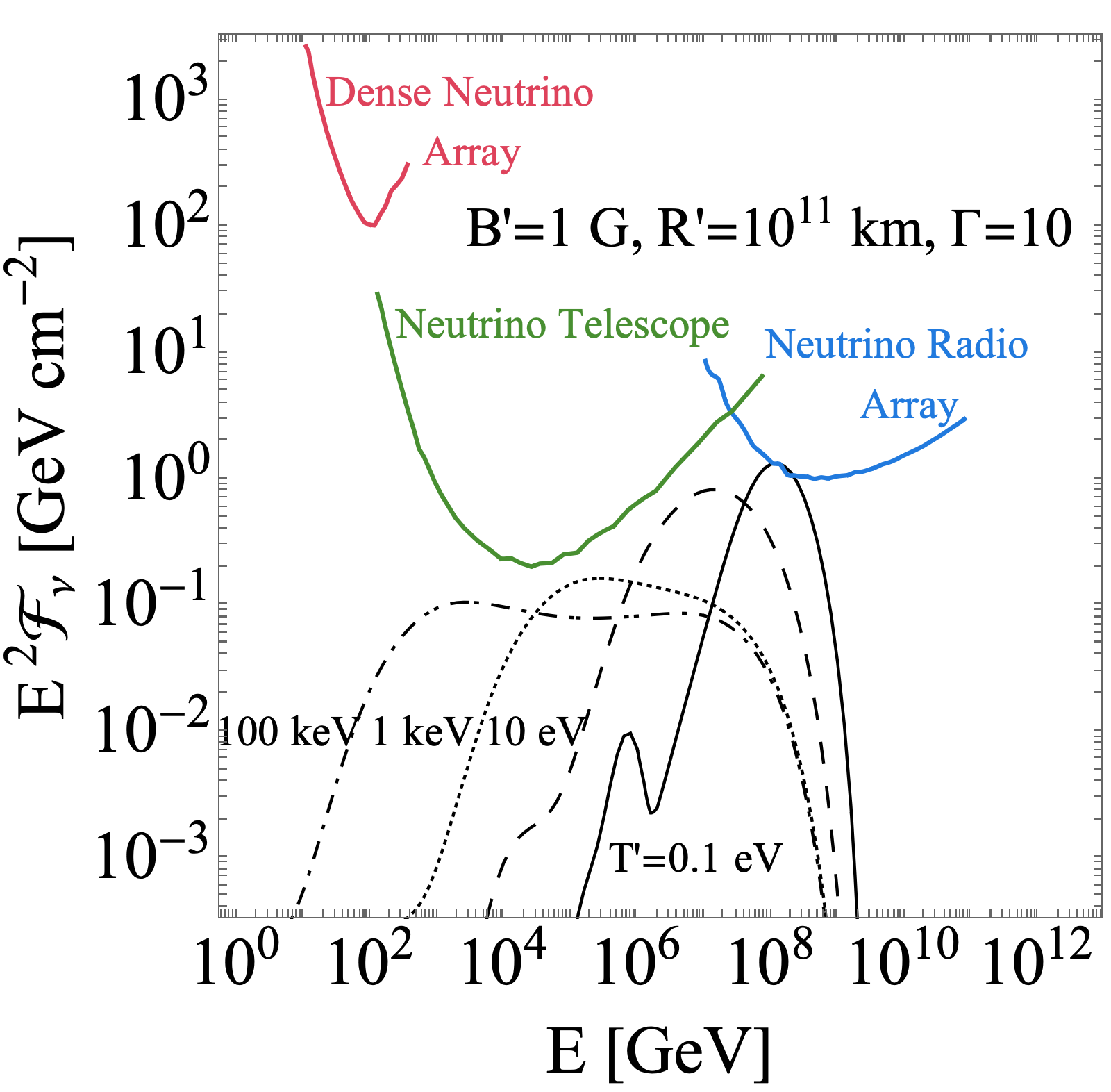}
    \caption{Experimental sensitivities for different model parameters. The figure shows the differential limits (colored curves) and the sensitivities to certain benchmark fluences (black curves) as a function of (observed) neutrino energy, where the curves refer to all flavors. 
    The different linestyles are referred to different values for the target photon temperature: all the other astrophysical parameters $B'$, $R'$, and $\Gamma'$ are fixed to benchmark values indicated in the figure. The fluences are normalized so that the sum of the expected events at all three experiments together is 2.44. }
    \label{varyagn}
\end{figure}

\section{Experimental sensitivity to astrophysical sources}

The model introduced in the previous section allows to characterize a generic astrophysical source by four parameters only: the effective photon temperature $T'$, the magnetic field $B'$, the source size $R'$, and the Doppler factor of expansion $\Gamma$. As a first application of the model, we now study how the sensitivity for detection at neutrino detectors with different energy ranges depends upon these parameters. We consider three classes of neutrino detectors, divided according to the energy range in which they are most sensitive. For each class we identify a benchmark representative experiment:
\begin{itemize} 
\item \textit{Dense neutrino arrays}, between $1$~GeV and $10^5$~GeV (e.g. PINGU, ORCA, DeepCore);
\item \textit{Neutrino telescopes}, between $10^5$~GeV and $10^7$~GeV (e.g. IceCube, KM3NeT, Antares); 
\item \textit{Neutrino radio arrays}, between $10^7$~GeV and $10^{12}$~GeV (e.g. ARIANNA, IceCube-Gen2 Radio Array, GRAND).
\end{itemize}
We choose as representatives DeepCore~\cite{Schulz:2009zz} for dense neutrino arrays, KM3NeT~\cite{Kappes:2007ci} for neutrino telescopes, and IceCube-Gen2 Radio Array~\cite{Blaufuss:2015muc} for neutrino radio arrays, respectively.

Before presenting a systematic scan of the parameter space of astrophysical sources, we show for a specific choice of parameters how the neutrino flux from the source compares with the sensitivities for neutrino detection in Fig.~\ref{varyagn}. The black curves are the all-flavor neutrino fluences for varying effective temperature of the target photons: they are normalized so that the sum of the events expected at all three experiments are equal to 2.44, corresponding to the background-free Feldman-Cousins 90\% sensitivity limit~\cite{Feldman:1997qc}. Since we impose this condition on the fluence, the result is independent of the observation time over which the detection is conducted. Qualitatively, an experiment is most suitable for detection when its energy range of sensitivity coincides with the energy range of the neutrinos produced in the source. Thus, lowering the photon temperature leads to spectra which are more peaked near their maximal energies and more easily detected by neutrino radio arrays. At higher temperatures, instead, the spectrum extends over decades of energy as a flat spectrum, and is more easily detected in the range below $1$ PeV by neutrino telescopes.

\begin{figure}[t]
    \centering
   \includegraphics[width=0.3\textwidth]{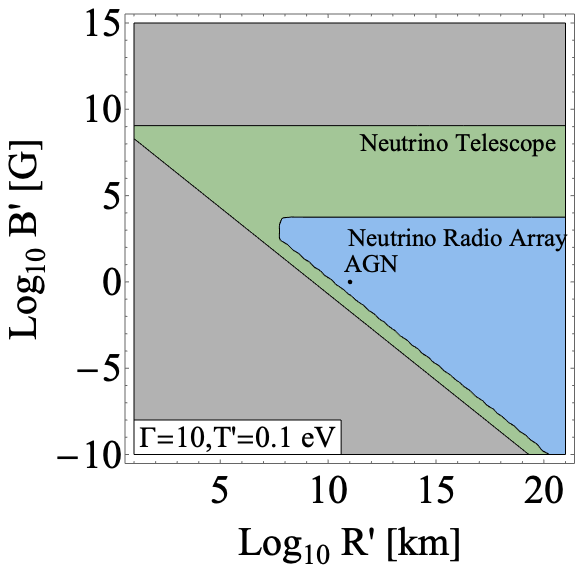}
   \includegraphics[width=0.3\textwidth]{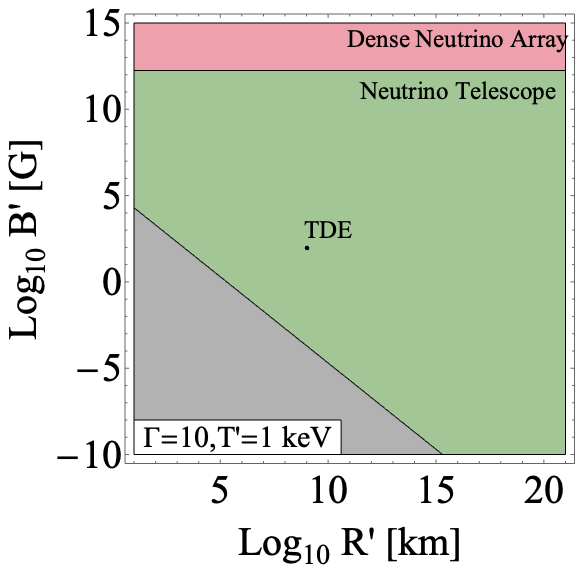}
   \includegraphics[width=0.3\textwidth]{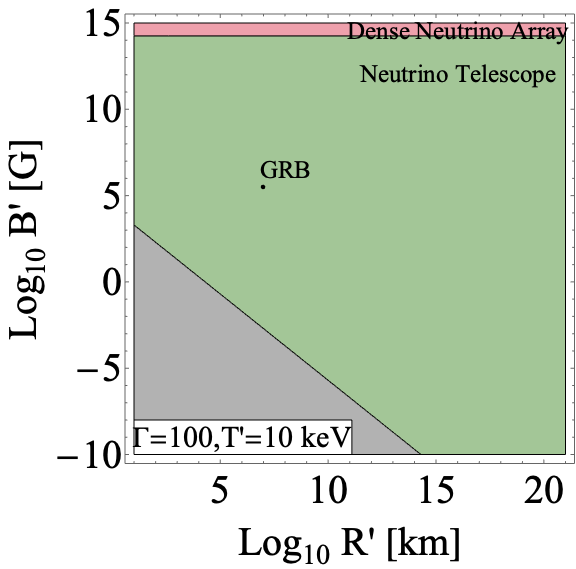}
    \caption{Determination of the most suitable experiment for the detection of astrophysical sources with parameters $\Gamma$, $T'$, $B'$ and $R'$. We show the Hillas plane divided according to the most suitable experiment for detection: the red, green and blue regions correspond, respectively, to our benchmark Dense Neutrino Array, Neutrino Telescope and Neutrino Radio Array as most sensitive experiments. The three panels correspond to different values for the effective temperature and Doppler factor. Typical values of $B'$, $R'$, $T'$ and $\Gamma$ for AGN, TDE and GRB sources are identified (the sources are identified by the order of magnitude of the effective temperature rather than by the precise value). In the gray regions, neutrino production is inefficient, because the maximal proton energy is below the threshold for $p\gamma$ interaction for all target photons.}
    \label{resulsensit}
\end{figure}

To quantify our comments on which experiment is more suitable for detection, for each experiment and for a given astrophysical source we normalize the (all-flavor) neutrino fluence $\mathcal{F}_\nu$ in such a way that  the expected number of events is 2.44.   For the fluence normalized in this way, we determine the total energy fluence as
\begin{equation} \label{eq:defxi}
    \xi=\int E \,  \mathcal{F}_\nu \text{d} E \, .
\end{equation}
This is the minimum energy fluence that must be emitted in neutrinos by the source in order to be detected by the experiment under examination. Therefore, the smaller $\xi$, the more suitable is the experiment for the detection of the source, since less energy needs be injected for detection. We refer the reader to Ref.~\cite{Fiorillo:2021hty} for a more detailed discussion on how $\xi$ is connected with the baryonic loading and the pion-production efficiency of the source. This criterion allows us to classify the parameter space according to which experiment is more suitable for detection: thus for each combination of the four parameters $T'$, $B'$, $R'$, and $\Gamma$, we determine $\xi$ for all three experiment classes and we choose the experiment class with the lowest value of $\xi$ as the one most suitable for detection. We show this classification for three benchmark choices of $T'$ and $\Gamma'$ in the Hillas plane, namely the $B'-R'$ plane, in Fig.~\ref{resulsensit}. These results show that detection with neutrino radio arrays is generally associated with low temperatures and high Doppler factors: in particular, we find that AGN are most easily detected by neutrino radio arrays, while GRBs and TDEs are most easily detected by neutrino telescopes.

\section{Flavor structure}

\begin{figure}[t!]
    \centering
    \includegraphics[width=0.3\textwidth]{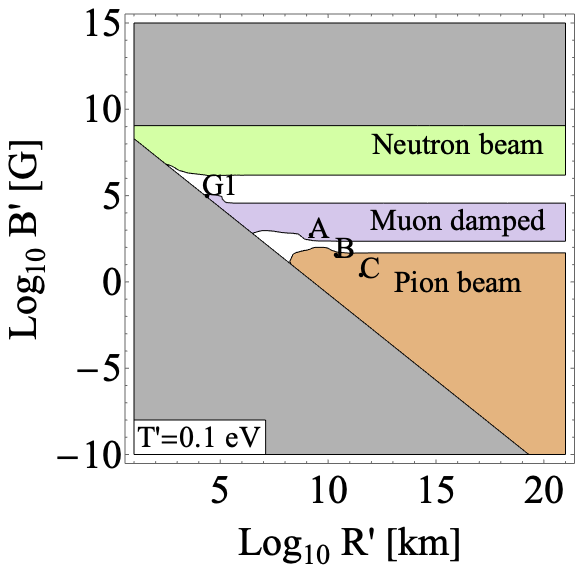}
    \includegraphics[width=0.3\textwidth]{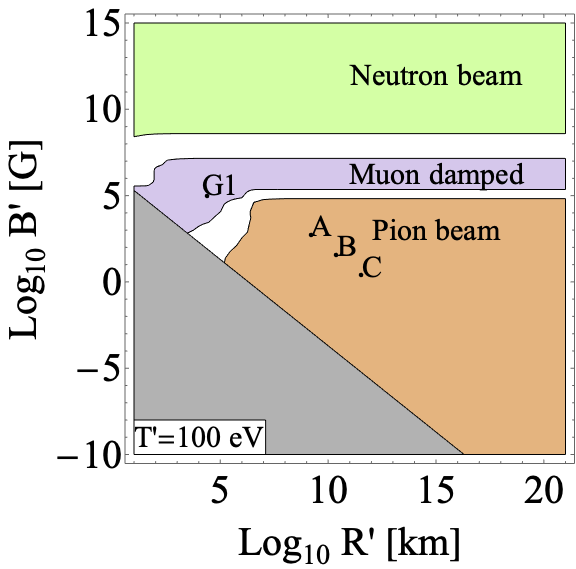}
 \includegraphics[width=0.3\textwidth]{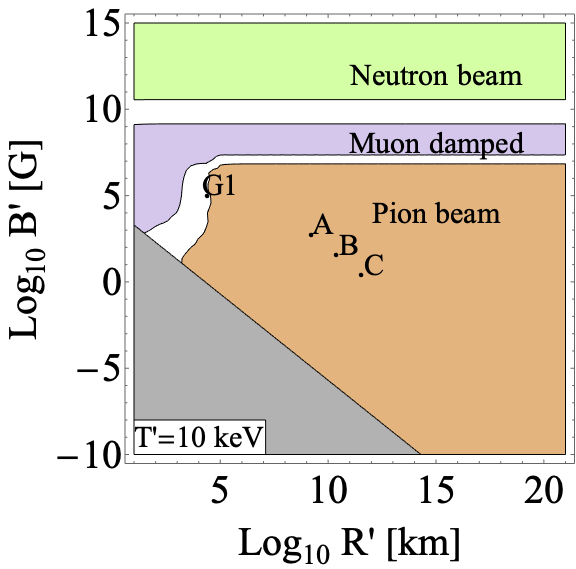}
    \caption{Flavor structure at the peak of the spectrum as a function of $R'$ and $B'$ (Hillas plot). We divide the Hillas plane according to the flavor composition at the peak of the spectrum: the neutron-beam region has a flavor ratio (at the source) between $(0.9:0.1:0)$ and $(1:0:0)$; the muon-damped region has a flavor ratio between $(0.1:0.9:0)$ and $(0:1:0)$; the pion-beam region has a flavor ratio between $(0.31:0.69:0)$ and $(0.36:0.64:0)$. In the white regions the flavor composition does not belong to any of the previous regimes. The three panels correspond to $T'=0.1$~eV, $T'=100$~eV and $T'=10$~keV. The gray regions correspond to inefficient pion production. Test points are indicated by A, B and C and by G1.}
    \label{flavorratios}
\end{figure}

\begin{figure}[t!]
    \centering
    \includegraphics[width=0.32\textwidth]{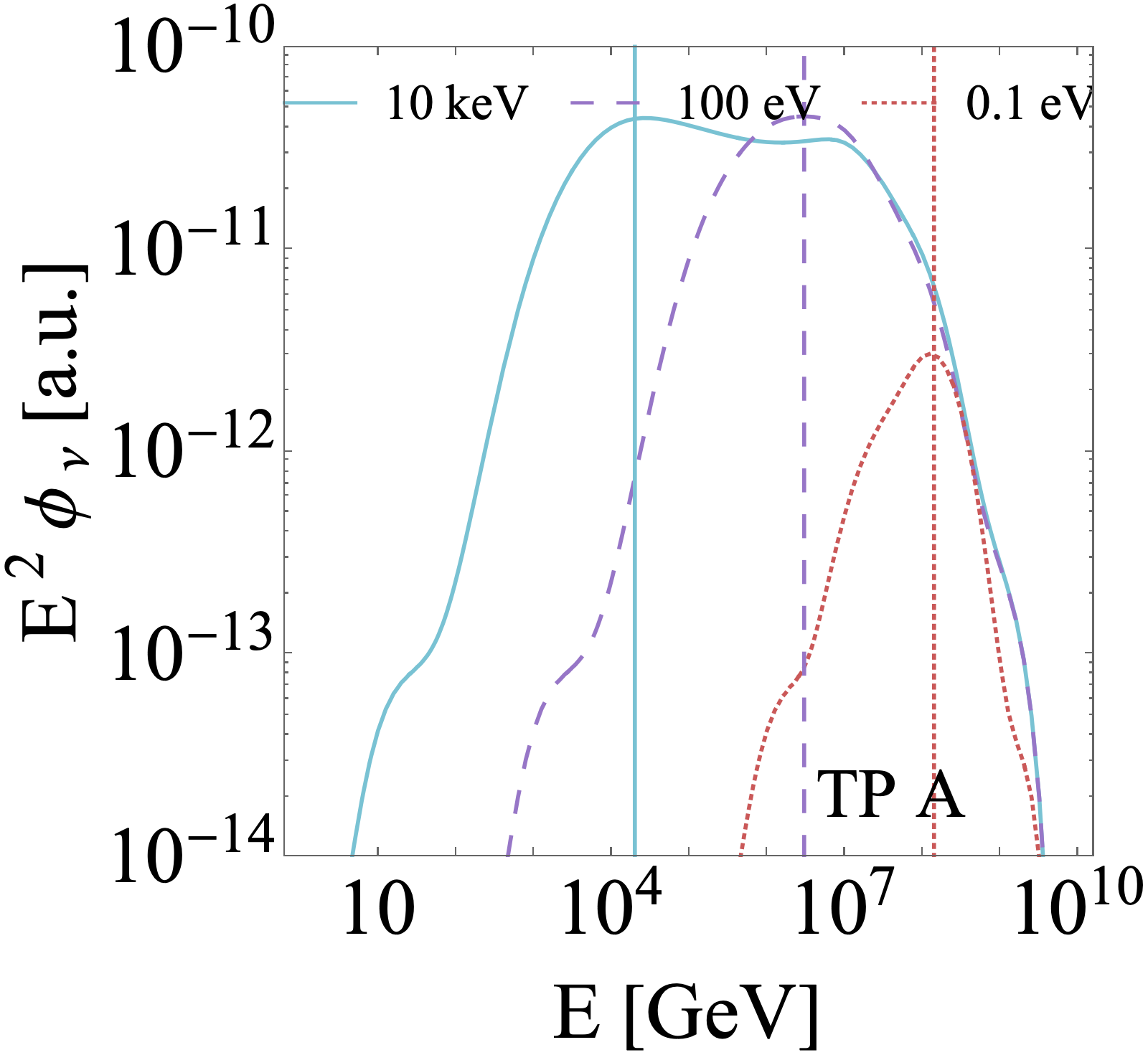}
    \includegraphics[width=0.3\textwidth]{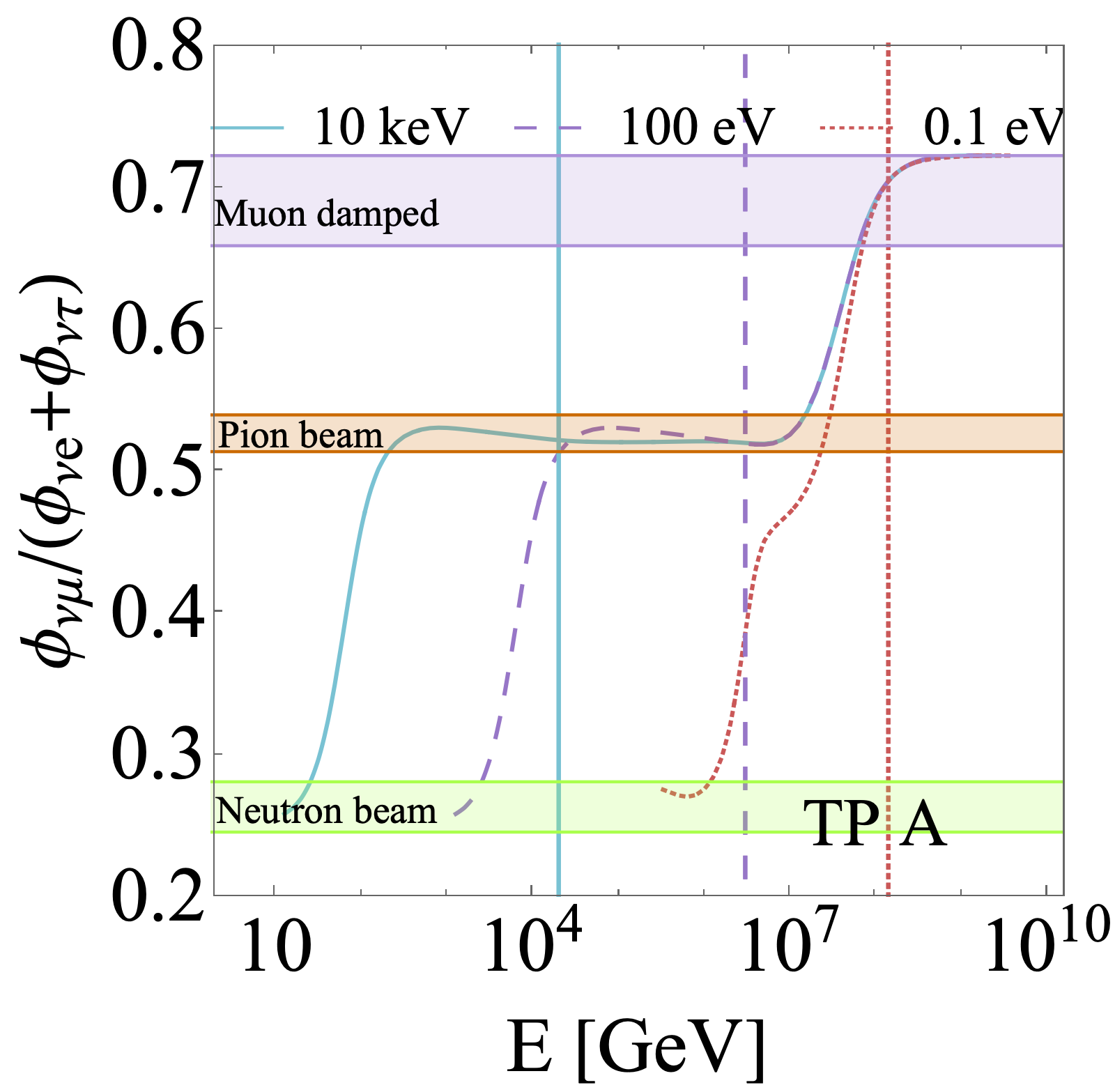}
    \caption{Neutrino fluxes and flavor ratios at Earth as a function of energy. In the left panel we show the all-flavor neutrino fluxes as a function of the energy for the TPA in Fig.~\ref{flavorratios}, which are chosen to simulate lower magnetic fields from left to right. The three curves to $T'=0.1$~eV, $T'=100$~eV and $T'=10$~keV; in all cases $\Gamma=10$. In the right panel we show the ratio between the muon and the sum of electron and tau neutrino and antineutrino differential flux as a function of the energy. The curves are represented only in the region in which the flux is at least $1/1000$ of its peak value. The horizontal bands identify the different flavor regime according to the quantitative criterion defined in the caption of Fig.~\ref{flavorratios}. In both panels we identify the peak energies of the neutrino fluxes with vertical lines: each color corresponds to the effective temperature according to the legend.} 
    \label{testpointse}
\end{figure}

The flavor structure of the neutrino flux is influenced by the target photon spectrum and by the astrophysical parameters of the source. In order to quantify this dependence, we classify the parameter space according to the flavor composition at the peak energy of the flux $E^2 \phi_\nu$: the flavor compositions we consider are pion beam, muon damped, and neutron beam. We represent in Fig.~\ref{flavorratios} the Hillas plane for three benchmark choices of temperatures, classified in different regions with this criterion. The flavor ratio at the peak of the spectrum is mainly influenced by the magnetic field: for low magnetic fields the full decay chain of pions and muons is active, leading to a pion-beam composition; raising the magnetic fields, muons are damped by synchrotron losses, leading to the muon-damped regime. Finally, for very large magnetic fields, synchrotron losses completely damp the spectrum and neutrinos from neutron decay dominate the spectrum, leading to a neutron beam composition. 

Our observations were based on the composition at the peak of the spectrum: however, an interesting feature is the energy dependence of the flavor composition. In Fig.~\ref{testpointse} we show this energy dependence for the test point A (TPA) identified in Fig.~\ref{flavorratios} and for varying effective temperature: for all the other test points the corresponding result is shown in Ref.~\cite{Fiorillo:2021hty}. We quantify the flavor composition by the ratio between the muon neutrino flux and the electron and tau neutrino flux $\phi_{\nu\mu}/(\phi_{\nu e}+\phi_{\nu\tau})$. The flavor composition exhibits marked transitions between different regimes, identified by horizontal bands in Fig.~\ref{testpointse}; it passes from neutron beam to pion beam to muon-damped regime for increasing energies. Decreasing the effective temperature pushes the peaks of the spectra to higher neutrino energies, as shown by the left panel of Fig.~\ref{testpointse}. Therefore the neutron-beam region moves to higher energies for lower effective temperatures. In addition, the peak of the spectrum moves progressively towards the region of muon damping.

\section{Conclusions}

Flexible models with few parameters are increasingly needed in order to efficiently test the origin of astrophysical neutrinos. Here we have proposed a unified model which can describe neutrinos produced via $p\gamma$ interactions from a generic astrophysical source in terms of a few parameters only, namely magnetic field, source size, Doppler factor, and effective photon temperature. After testing the model on neutrino production in AGN and GRBs, which is well reproduced, we have applied it to determine both the experimental sensitivity of different neutrino detectors and the flavor composition as a function of the parameters of the astrophysical sources. However, the model can be applied in a number of different contexts: in particular, it provides an easy way of parameterizing the astrophysical neutrino production in systematic searches for neutrino sources (e.g. multiplet and stacking analyses) and in studies of Beyond Standard Model physics effects on high-energy neutrinos.

\acknowledgments{This work was supported by the Italian grant 2017W4HA7S “NAT-NET: Neutrino and As- troparticle Theory Network” (PRIN 2017) funded by the Italian Ministero dell’Istruzione, dell’Universit\'a e della Ricerca (MIUR), and Iniziativa Specifica TAsP of INFN.}

\end{document}